\documentstyle[preprint,eqsecnum,aps]{revtex}
\tighten
\begin{document}
\input FEYNMAN
\title{Bound on nonlocal scale from $g-2$ of muon\\
in a nonlocal WS model}
\author{Gaitri Saini\thanks{supported by University Grants
Commision(UGC),India;
e-mail:ocrgs@iitk.ernet.in} and Satish D.Joglekar\thanks{e-mail:sdj@iitk.
ernet.in}}
\address{Department of Physics\\ I.I.T Kanpur\\ Kanpur,208016,India}
\date{\today}
\maketitle

\begin{abstract}

We consider the nonlocal version of the Weinberg Salam model
(following Kleppe et al.\ ) with a finite parameter $\Lambda$
signifying a fundamental length scale.We calculate the extra
contributions to the anomalous magnetic moment of the muon
coming from the nonlocal structure in this model.We find that
the nonlocal  contribution can be comparable to weak contributions
and goes to increase theoretical estimates.We use this calculation
to determine
the limit on the scale of nonlocality.We obtain the result $1/\Lambda
\stackrel{<}{\sim} 3 \times 10^{-16} $cm,which could be improved 
when present experimental errors narrow down.
\end{abstract}
\vspace{50 mm}
\pagebreak
\section{Introduction}

The presently sucessful theories of strong weak and electromagnetic 
interactions are all local quantum field theories: they assume a local
interaction. One of the features of local QFTs is the presence of ultraviolet
divergences that arise in the large momentum region in the loop integrations 
or equivalently, at short distances. Operator products of fields at the same 
$x$ are ill-defined due to such ultraviolet divergences. A natural procedure
for avoiding such ultraviolet divergences has been to introduce a momentum
cut-off $\Lambda$ or equivalently a small space-time separation 
$\epsilon_{\mu} (\sim \frac{1}{\Lambda})$ into the theory. Until recently,
all attempts to introduce such cutoffs into the local QFT led to theories
that were unitary only as $\Lambda \rightarrow \infty$. Hence such procedures
were used only as regularizations: a physically meaningful theory required
that renormalization be carried out and then $\Lambda$ let go to 
infinity. Hence 
such a introduction of a cut off was looked upon as a mathematical necessity
only, without a direct physical significance. 

Recently, a way of regularizing quantum field theories  was introduced
~\cite{emkw} and developed further ~\cite{kw,kws} by the 
introduction of a cut off
parameter $\Lambda$ in such a manner that (i) the theory is
unitary for any finite
$\Lambda$, (ii) all continuous symmetries of the theory are preserved 
in some
(altered) form. Such theories necessarily contain derivatives of fields
to all orders and hence are essentially nonlocal. Such a regularization has 
been labelled ``nonlocal regularization''.

Since the  theories are physically meaningful even for finite $\Lambda$, 
it has
been suggested by Kleppe and Woodard ~\cite{kw}, that one could look upon
the regularized theory, containing the parameter $\Lambda$, as the actual
physical theory with an inherent  distance scale $\sim\frac{1}{\Lambda}$
in it. Thus, according to this viewpoint, the nonlocal version of the theory
is not a regularization ( i.e. a mere mathematical necessity), but a
physical action that could embody a granular structure of space-time through
a parameter $\Lambda$.

A natural question one can ask at this stage is what are the experimental
limits on this parameter $\Lambda$. From this viewpoint, we consider using
the anomalous 
magnetic moment$ (g-2)$ of the muon,~\cite{fp}; which is known to one 
part in $10^9$. The contributions to $(g-2)$ in the local theory 
(standard model) have been well discussed  and accurately calculated
~\cite{kin} and are known to agree very well with experiment\cite{fp}.

It is the uncertainities in the experimental and theoretical values of 
$(g-2)$ that can set an upper limit on the nonlocal contribution to $(g-2)$.
From the theoretically calculated contribution to $(g-2)$, one can then set
a lower bound on the nonlocality parameter $\Lambda$. This is the purpose of
the present work.

One can also expect nonlocality to make much difference in the operator 
product expansion at short distances, when applied to deep inelastic 
scattering and electroproduction. We expect to report it elsewhere. 

While we can look upon this calculation as a limit on the scale of 
granularity of space time, alternative interpretations of the result
are also possible. For example, muon may not be a point particle as
assumed in the local theory, but may have an internal structure, which
may directly introduce nonlocality in its interaction with electromagnetic
and weak bosons. In such a case, if we assume that our nonlocal theory 
is an effective theory, these results tell us about the scale of 
compositeness of the muon. Now such calculations in composite models have
been performed in literature ~\cite{brod,pec}; these calculations
 however,necessarily require a number of adhoc assumptions
  in order to perform them.
 The calculations presented here on the other hand,
are based on the self consistent unitary BRS invariant model involving
the scale $\Lambda$ and does not involve ad hoc procedures. Of course, our
procedure while it gives a mathematically rigorous way of calculating nonlocal
contributions to $(g-2)$,  cannot shed light on how the compositeness scale
$\Lambda$ has arisen.

We should also mention that calculation of fundamental scale of space-time
has been attempted in a similar manner in another formulation of nonlocal
quantum field theory, viz. the stochastic quantization method, and 
 similar results have been obtained ~\cite{nam}. However, 
 these nonlocal methods are
much more cumbersome and require extra regularizations in addition to the
introduction of nonlocality. The formulation used in our work ~\cite{kw} is
neat and lends easily to calculations. Further, the calculation done for the 
stochastic formulation is for the photon exchange diagram only. In our
formulation, we on the other hand calculate all contributions including 
W, Z exchanges. While the latter contributions in local theory are small
compared to the $\gamma$-exchange diagram the nonlocal contributions from
all the $\gamma$-exchange, W-exchange and Z-exchange 
(and the related diagrams) are
of the same order [i.e. $O(m^2/\Lambda^2)$ ].

\section{The method of nonlocalization}
\label{tmon}

In this section, we briefly review the method of nonlocalization~\cite{emkw,kw}
for obtaining a nonlocal theory from a local one. The method has been discussed
extensively in many of the works on nonlocal 
regularization~\cite{kw,kws,cdm,par}  but we 
recapitulate the salient features here and also introduce our notations and 
conventions. The first step is the nonlocalization of the local action using
'smearing operators' and is applicable to any local theory which can be 
formulated perturbatively. With this construct, for every local symmetry of 
the local action there is a corresponding nonlocal symmetry for the nonlocal
action. This symmetry reduces to the local symmetry in the local limit
of the nonlocal theory ~\cite{kw}. The classical nonlocal theory 
is quantized using the
functional formalism, and in order that the quantized theory respects the 
nonlocal symmetry it is necessary that there should (except for anomalous
theories) exist a measure factor which makes the path integral measure
invariant under the nonlocal symmetry. For nonabelian gauge theories
for example, the required measure factor is nontrivial. The second step in
nonlocalization is the construction of such a measure factor such that
the resultant theory is finite, Poincare invariant and perturbatively unitary.
In section~\ref{phi4} we outline the method of nonlocalization 
using ${\phi}^4$ theory
as an example. In section~\ref{nl-ws} we nonlocalize the 
Weinberg Salam model, gauge 
fixed in the Feynman gauge and write down the Feynman rules needed to
evaluate the anomalous magnetic moment of the muon in the nonlocal theory.

\subsection{Nonlocal ${\phi}^4$ theory}  
\label{phi4}

Consider a local theory with the action written as a sum of the free and 
interacting parts 
\begin{equation}
\label{locth}
S[\phi] = \frac{1}{2} \sum_{\phi} \int d^{4}x \phi {\cal F} \phi + 
          I[\phi]\; ,
\end{equation}
where $\phi$ represents the fields (fermionic, bosonic) of the theory with the
appropriate spacetime and internal symmetry group indices. ${\cal F}$ is 
the 'kinetic' operator for the field $\phi$ and $I[\phi]$ is the 
interaction part of $S[\phi]$. For the ${\phi}^4$ theory,
\begin{equation}\label{f-phi4}
{\cal F} = (-{\partial}^2 - {\mu}^2)
\end{equation}
and
\begin{equation}\label{i-phi4}
I[\phi] = - \frac{\lambda}{4!} {\phi}^4
\end{equation}
Nonlocalization of $S[\phi]$ is carried out using a 'smearing' operator 
defined in terms of the kinetic operator ${\cal F}$ of the theory as
\begin{equation}\label{sm-op}
{\cal E} \equiv  exp[\frac{{\cal F}}{2 \Lambda^2}]. 
\end{equation}
$\Lambda$ is the scale of nonlocality. With the help of the smearing operator 
${\cal E}$ we define a smeared field $\hat{\phi}$ by
\begin{equation}\label{phi-hat}
 \hat{\phi}= {\cal E}^{-1} \phi
\end{equation}
Next, for every field $\phi$ we introduce an auxillary 'shadow' field 
${\phi}^{sh}$ of the same type as $\phi$ which couples to $\phi$ through
an auxillary action ${\cal S}$ given by
\begin{equation}\label{sh-action}
{\cal S}[{\phi},{\phi}^{sh}] \equiv \frac{1}{2} \sum_{\phi}
    \int d^4 x\hat{\phi}{\cal F} 
    \hat{\phi} 
    - \frac{1}{2} \sum_{{\phi}^{sh}} \int d^4x{\phi}^{sh}
    {\cal O}^{-1} {\phi}^{sh} 
    + I[ \phi + {\phi}^{sh}]
\end{equation}
where ${\cal O}$ is the 'shadow' kinetic operator defined as 
\begin{equation}\label{shadow-op}
{\cal O} \equiv ({\cal E}^2 - 1) {\cal F}^{-1}
\end{equation}
The action for the nonlocalized theory $\hat{S}[\phi]$ is then 
defined by
\begin{equation}  \label{nl-action}
\hat{S}[\phi] \equiv {\cal S} [ \phi\; , \phi^{sh}(\phi)]
\end{equation}
where ${\phi}^{sh}[\phi]$ is the solution of the classical shadow field
equation
\begin{equation}\label{cl-eq}
\frac{\delta {\cal S}}{\delta {\phi}^{sh}} [\phi\; ,{\phi}^{sh} ]
  = 0 
\end{equation}
Quantization is carried out in the path integral formulation. The quantization
rule is given by
\begin{equation}\label{path-int}
\langle T^*(O[\phi]) \rangle_{\cal E} \equiv \int D[\phi]\mu[\phi] 
     O[\hat{\phi}]  e^{i\hat{S}  [\phi] }
\end{equation}
Here $O$ is any operator taken as a functional of fields. $\mu[\phi]$ is the 
measure factor defined such that $D[\phi]\mu[\phi]$ is invariant under
the nonlocal generalization of the local symmetry. For the ${\phi}^4$ theory
$\mu[\phi] = 1$. For nonlocalized nonabelian gauge theories for example this
measure factor is nontrivial.

The Feynman rules for the nonlocal theory are simple extensions of the rules
of the local theory.
In the local theory we leave the local propagator given 
by\footnote{In this and the following sections the Feynman rules
are those of a Minkowski space formulation.The loop integrations
are well defined only in Euclidean space, however,and we evaluate
them by a formal Wick rotation\cite{emkw}.}
\begin{equation}\label{loc-prop}
\frac{i}{{\cal F} + i\epsilon} 
=-i\int_0^\infty \frac{d\tau}{\Lambda^2} e^{\frac{\tau{\cal F}}{\Lambda^2}}
\end{equation}
In the nonlocal formulation there are two kinds of propagators,
 'smeared' or an 'unbarred' propagator
\begin{equation}\label{smeared-prop}
\frac{i{\cal E}^2}{{\cal F} + i\epsilon} 
=-i\int_1^\infty \frac{d\tau}{\Lambda^2} e^{\frac{\tau{\cal F}}{\Lambda^2}}
\end{equation}
and a 'barred' or shadow propagator
\begin{equation}\label{barred-prop}
\frac{i(1 - {\cal E}^2)}{{\cal F}} = - i {\cal O}
=-i\int_0^1\frac{d\tau}{\Lambda^2} e^{\frac{\tau{\cal F}}{\Lambda^2}}
\end{equation}
For $\lambda {\phi}^4$ theory, these are represented 
graphically in Fig.~\ref{phi4prop}.
The vertices are the same as those of the local theory except that in the
nonlocal theory we have additional vertices coming from the measure factor,
whenever this factor is different from unity. For computing Feynman diagrams
in the nonlocal theory the following points are to be noted.

a) the external lines of a given diagram can only be 'smeared' lines

b) The symmetry factor for any diagram is computed without distinguishing
between barred and unbarred lines.

c) The loop integrations are well defined in the Euclidean space because of
the exponential damping factors coming from propagators within loops

d) The internal lines within a loop can either be smeared or barred. We sum over
all possibilities excluding the diagrams with loop(s) made up entirely of
'barred' lines since shadow fields by construct obey Eq.~\ref{cl-eq} and we
do not functionally integrate over them. Note also that including 'shadow loops'
would give a theory which in effect is the same as the unregulated local theory.
This is clarified by the following example:
Consider the tadpole self energy diagram in nonlocal ${\phi}^4$ theory 
(Fig.~\ref{uloop}) which using Eqs. ~\ref{smeared-prop} and~\ref{f-phi4}  
is given as\footnote
{The notations for the Feynman rules for the local theory are those
given in Appendix(B) of Ref.~\cite{ch-li}}.
\begin{equation}\label{sm-tad}
 = (-i) (\frac{-i\lambda}{2}) \int \frac{d^4l}{(2\pi)^4}
   \int_1^\infty\frac{d\tau}{\Lambda^2}e^{\frac{\tau}{\Lambda^2} ( l^2 
   - {\mu}^2)}
\end{equation}
the vertex factor is $(- i \lambda )$ and l is the loop momentum.
The shadow loop self energy diagram shown in Fig.~\ref{bloop} is given by
(using Eqs.~\ref{f-phi4} and ~\ref{barred-prop}):
\begin{equation}\label{sh-tad}
 = (-i) (\frac{-i\lambda}{2}) \int \frac{d^4l}{(2\pi)^4}
   \int_0^1\frac{d\tau}{\Lambda^2}e^{\frac{\tau}{\Lambda^2} ( l^2 
   - {\mu}^2)}
\end{equation}
If we take into consideration the 'shadow loop' also, we would obtain
,upon adding Eqs.~\ref{sm-tad} and ~\ref{sh-tad}
\begin{equation}\label{loc-tad}
\frac{i \lambda}{2} \int \frac{d^4l}{(2\pi)^4} 
\frac{1}{ l^2 -{\mu}^2 + i\epsilon}
\end{equation}
which is the tadpole diagram of the local theory. 
Note too that the shadow loop diagram here is ill-defined. In fact this example
also shows how the finiteness of the nonlocalized theory arises. The 
divergences in the diagrams of the local theory can be seen as arising from
the inclusion of the region around the origin of integration in the 
Schwinger parameter ($\tau$) space (see Eq.~\ref{loc-prop}) 
The nonlocalized theory is finite because of the exclusion of the unit 
hypercube at origin from the region of integration in parameter space.
This is effected by the exclusion of shadow loops. 
\subsection{Nonlocal WS model in Feynman gauge}\label{nl-ws}
In this section we nonlocalize the WS model which has been gauge fixed in the
Feynman gauge. In this paper since we are interested in 
evaluating the one loop nonlocal electroweak contribution to the magnetic
moment anomaly of the muon, $ a_{\mu} $, we will write down explicitly
only the Feynman rules necessary for the calculation of $a_{\mu}$. However,
it is understood that we are nonlocalizing the full local theory and the 
complete set of Feynman rules are given accordingly, as outlined in 
section~\ref{phi4}.
In order to introduce the notations and definitions we first write down
the action for the local WS model~\cite{ch-li} in the $R_\xi$ gauge, as a 
sum of the free and interacting parts:
\begin{equation}
S=\int d^4x{\cal L}
\end{equation}
where
\begin{eqnarray*}
{\cal L}&=&\frac{1}{2}\sum_{+,-}W_{\mu}{\cal F}_{\pm}^{\mu\nu}W_{\nu}\;+\;
\frac{1}{2}Z_{\mu}{\cal F}_{Z}^{\mu\nu}Z_{\nu}\; + \; 
\frac{1}{2}A_{\mu}{\cal F}_{A}^{\mu\nu}A_{\nu} \\
& &+\frac{1}{2}\sum_{+,-,1,2}\phi_i{\cal F}^i\phi_i \; + \; \mu^ -(i\not 
\partial-m)
\mu^-  \;  +\; \cdots \\
 & &
 +\; I[W^{\pm},A,Z,\phi^{\pm},\phi_1,\phi_2,ghosts,leptons,quarks]
\end{eqnarray*}
The dots in the above equation stand for the kinetic terms for the other 
leptons, quarks, ghosts. Here$W^{\pm}$,$Z_{\mu}$,$A_{\mu}$ are the vector 
bosons,
 $\phi^{\pm}$ and $\phi_2$ are the three would be goldstone bosons and
 $\phi_1$ is the Higg's scalar.
I represents the interaction part of the lagrangian.

The kinetic operators for the fields shown in the above equation are:
\begin{eqnarray*}
{\cal F}^{\mu\nu}_{\pm}&=&(\partial^2\; +\; m_W^2)g^{\mu\nu}\; +\; 
(1/\xi-1)\partial^{\mu}\partial^{\nu} \\
{\cal F}^{\mu\nu}_{z}&=&(\partial^2\ +\ m_z^2)g^{\mu\nu}\; +\; 
(1/\xi-1)\partial^{\mu}\partial^{\nu} \\
{\cal F}^{\mu\nu}_{A}&=&\partial^2g^{\mu\nu}\; +\; 
(1/\xi-1)\partial^{\mu}\partial^{\nu}
\end{eqnarray*}
\begin{eqnarray*}
{\cal F}_{\phi^{\pm}}&=&-\partial^2\; -\; \xi m_W^2 \\
{\cal F}_{\phi_2}&=&-\partial^2\; -\; \xi m_z^2 \\
{\cal F}_{\phi_1}&=&-\partial^2\; -\; 2\mu^2\; \; (\mu^2>0)
\end{eqnarray*}
We now nonlocalize the theory defined above, according to the procedure 
outlined in section~\ref{phi4}. For simplicity we consider the theory in the
Feynman gauge ($\xi=1$). The smearing operators for all bosonic fields are given
by Eq.~\ref{sm-op} where the kinetic operators ${\cal F}$ are defined
in the equations above with $\xi=1$. For the fermions, it is simplest to define the smearing
operator as a scalar operator
\begin{equation}\label{ferm-op}
{\cal E}_{\psi}=e^{-\partial^2-\mu^2}.
\end{equation}
Having defined the smearing operators for all fields we define smeared
fields according to Eq.~\ref{phi-hat}  and also introduce shadow fields which 
couple to the smeared fields via the auxillary action given by 
Eq.~\ref{sh-action}
The nonlocal action is finally defined using Eq.~\ref{nl-action}. 
We assume that the
measure factor which make the quantized theory perturbatively unitary and 
Poincare invariant, exists. However, we do not discuss it in this paper at all
since its vertices are not needed in the evaluation of $a_\mu$ 
to one loop order in the 
nonlocal theory.(The measure factor vertices are necessarily of order
$\hbar$ and have external gauge boson lines only, and
hence they can contribute to $a_{\mu}$ only at two and higher loop
 order.).The Feynman rules needed for computing $a_{\mu}$ are as follows.
The propagators are shown in Fig~\ref{props}. The vertex factors are the
same as those of the local theory and can be found in Appendix B of 
Ref.~\cite{ch-li} 
Of course, it must be borne in mind that the vertices connect smeared and/or
shadow lines and not local ones. 
\section{ Calculation of nonlocal contribution to $ {\em a}_{\mu}$}
\label{3}
In this section we evaluate the one-loop
nonlocal corrections to $a_{\mu}$ arising in the nonlocal electroweak
theory discussed in Section~\ref{nl-ws}.The leading order corrections 
are of order $\alpha m^2/\Lambda^2$ where $\alpha$ is the fine structure
constant and $m$ the muon mass.We assume that the scale of nonlocality
$\Lambda\gg M_W$,and therefore 
neglect corrections of order$(\Lambda^2)^{-n},n>1$.

\subsection{Feynman diagrams contributing}
\label{3a}
The one loop electromagnetic contributions to $a_{\mu}$ in the nonlocal
theory will come from the diagrams in Fig.~\ref{qedloops} where 
the unbarred (barred)
lines are the smeared (shadow) muon and photon lines. External lines are
always smeared lines as has been pointed out in Section~\ref{phi4}.
For compactnes of notation we represent the  diagrams of 
Fig.\ref{qedloops} as shown in
Fig.~\ref{fvert}
The term 'barred variations' stands for all diagrams obtained from the unbarred
diagram by replacing one or more of the internal lines with barred , i.e.,
shadow lines, {\em excluding} the case  where all the
lines in a loop are barred.

To calculate the one loop weak contributions we have to consider the 
diagrams shown in Fig.~\ref{sm-weak} along with the barred variations of each
diagram as explained above.
From these diagrams we are interested only in (on shell) contributions
proportional to $\bar{u}(q)\sigma_{\mu\nu}(p-q)^{\nu}u(p)$ 
(where $q$ and $p$ are the final and initial momenta, respectively) and will
ignore the rest of the terms.
Firstly,notice that the sum of all diagrams
given in Fig.~\ref{qedloops} for example, and the shadow loop diagram in 
Fig.~\ref{sh.qed}
gives the diagram of the local theory (Fig.~\ref{l.qed}).
(Refer also to the dicussion
at the end of section~\ref{phi4}  and Eqs.~\ref{sm-tad}-~\ref{loc-tad})
Now, the local contributions for $a_{\mu}$ coming from the diagram in 
Fig.~\ref{l.qed} of the local theory is finite. 
The nonlocal contributions come from
diagrams of Fig.~\ref{qedloops} are also finite\footnote{Since
 the nonlocal theory by construct is a finite theory ($\Lambda$
finite), all diagrams of the theory are finite. Further, for contributions
to $a_{\mu}$ there is no contribution proportional to $(\Lambda^2)^n,
n > 0$, integer. This is because in the limit $\Lambda\rightarrow \infty$,
the nonlocal theory reduces to the local theory and nonlocal contributions
to $a_{\mu}$ reduce to the (finite) local contributions. Therefore all 
nonlocal contributions will be propotional to $(\Lambda^2)^n$ where n is an
integer $\leq 0$.}
. Hence the contributions for $a_{\mu}$
from the shadow loop have to be finite.
\footnote{In general, the shadow loops are ill-defined whenever the 
corresponding
loops in the local theory are, and we cannot consider them for 
manipulations (see end of Section~\ref{phi4}  for an example).} 
Therefore instead of considering all the
diagrams of Fig.~\ref{qedloops} we need consider only the 
single shadow loop diagram of Fig.~\ref{sh.qed}. 
Then the contribution to $a_{\mu}$ in the nonlocal theory will
simply be the difference between the local and shadow loop contributions. 
Hence, the correction to $a_{\mu}$ due to nonlocality will be given by the 
negative of the shadow loop contribution.

The same is true for weak contributions. Therefore instead of considering the
diagrams
of Fig.~\ref{sm-weak} along with their barred variations  
we need consider only the
weak boson exchange shadow loop diagrams of Fig.~\ref{sh-weak} 
along with the shadow
diagram of Fig.~\ref{sh.qed} of $\gamma$ exchange.
We will evaluate these in the next section.

\subsection{Calculation for $a_{\mu}$}
\label{3b}   

In order to calculate the nonlocal contributions to $a_{\mu}$,
we have seen that we only have to calculate the shadow diagrams 
of Figs.~\ref{sh.qed} and ~\ref{sh-weak};and from these we need 
to extract the on-shell
contribution proportional to $\bar{u}(q)\sigma_{\rho\nu}u(p)$.
We will work  in the Feynman gauge ($\xi=1$)
.We shall be brief
in our discussions and only elaborate those points which are new to 
the reader because of the use of a nonlocal theory.

We shall ascribe a common momentum flow to all diagrams as shown
in Fig.~\ref{m-route}.

The $\gamma$-exchange shadow  diagram of Fig.~\ref{sh.qed} can be written as
\begin{eqnarray}
i\Delta^{(1)}\Gamma_{\rho}(p,q)&= &-e^3\int\frac{d^4l}{(2\pi)^4}
\int_{0}^{1}\frac{d\tau_1}{\Lambda^2}\int_{0}^{1}\frac{d\tau_2}
{\Lambda^{2}}\int_{0}^{1}\frac{d\tau_{3}}{\Lambda^2}
 \gamma_{\lambda}
(\not l-\not p+\not q+m)\gamma_{\rho}(\not l+m)
\gamma^{\lambda} \nonumber \\ &\times& exp\left[\frac{1}{\Lambda^{2}}
(\tau_{1}(l-p+q)^{2}+\tau_{2}l^{2}+
\tau_{3}(p-l)^{2}-(\tau_{1}+\tau_{2})m^{2})\right]
\label{3.1}
\end{eqnarray}

where an extra minus sign has been added because it is the -ve
of the shadow diagram that gives the nonlocal contribution to
$a_{\mu}$.The above expression is understood to be sandwiched between
$\bar{u}(q)$ and $ u(p)$.Then $\Delta^{(1)}\Gamma$ is a sum of $\gamma_{\rho}$
and $\sigma_{\rho\nu}(p-q)^{\nu}$type terms.
The following common features are noted.

(1).A shift of the momentum variable
\begin{eqnarray}\label{3.2}
l&\rightarrow l+\frac{\tau_{1}+\tau_{3}}{\tau}p-\frac{\tau_{1}}{\tau}q\;\; ;&
\;\;\;\;\;\;\tau\equiv\tau_{1}+\tau_{2}+\tau_{3}
\end{eqnarray}
removes from the exponent cross terms in l, and further,the change of 
variables above is commom to all diagrams in view of the common momentum 
routing in Fig.~\ref{m-route}.

(2).The exponent then assumes the form
\begin{eqnarray}
\lefteqn{exp\left[\frac{\tau l^2}{\Lambda^{2}}+
\frac{1}{\Lambda^{2}}f(m^{2},p,q;\tau_{i})\right]}
\nonumber \\
 & &=exp\left[\frac{\tau l^2}{\Lambda^{2}}\right]\left[1+\frac{1}
 {\Lambda^{2}}f+O(\frac{1}{\Lambda^{4}})+\cdots\right]
\label{3.3}
\end{eqnarray}
As we shall see, the first term in the above expansion itself 
contributes the leading nonlocal
correction of O($m^{2}/\Lambda^{2}$) and when this happens the contributions 
from further terms in the series (which are all well defined:
recall that $f_i$ are polynomials in $\tau_i$'s and the range of $\tau_i$'s
is 0 to 1,i.e,finite)only give nonleading contributions and 
hence can be dropped.

(3).Note next that in an expression such as 
\begin{equation}\label{3.4}
\int_0^1d\tau_1\int_0^1d\tau_2\int_0^1d\tau_3\;\; 
exp\left[\frac{\tau}{\Lambda^2}l^2\right]
h(p,q,\tau_i)
\end{equation}
as the integral ranges as well as the exponential are symmetric under 
simultaneous interchange of any two $\tau_i$'s,the h($\tau_i$)'s
can be symmetrized.

(4).We need only the following $l$ integral which can be done following 
a Wick rotation \cite{emkw}:
\begin{equation}\label{3.5}
\int\!\frac{d^4l}{(2\pi)^4}exp[\frac{\tau}{\Lambda^2}l^2]
=\frac{i}{16\pi^2}\frac{\Lambda^4}{\tau^2}
\end{equation}

(5).When sandwiched between $\bar{u}(q)$ and $u(p)$,on mass shell
for the muon, the following expressions contribute,as indicated below.
Here,we have used Gordon decomposition and dropped $\gamma_{\rho}$ type
terms(that contribute to the electric form factor)
\begin{eqnarray}
\gamma^{\rho}\not p \approx 0&;&\;\;\;\not q\gamma^{\rho}\approx 0 \nonumber \\
\gamma^{\rho}\not q\approx -2i\sigma^{\rho\nu}q_{\nu}&;&\;\;\;\not p
\gamma^{\rho}
\approx 2i\sigma^{\rho\nu}p_{\nu}\label{3.7} \\
\not p\gamma_{\rho}\not q&\approx&-2im\sigma_{\rho\eta}(q-p)^{\eta}\nonumber 
\end{eqnarray}

Using the statements made above,we can evaluate $\Delta^{(1)}\Gamma_{\rho}$
of Eq.(\ref{3.1})and we find
\begin{eqnarray}
\lefteqn{i\Delta^{(1)}\Gamma_{\rho}(p,q)=(\frac{e}{2m}
\sigma_{\rho\mu}(p-q)^{\mu}
)\frac{e^2}{16\pi^2}\frac{m^2}{\Lambda^2}\times(-8)}\nonumber \\ &\;&\times
\int_0^1d\tau_1\int_0^1d\tau_2\int_0^1d\tau_3\frac{1}{\tau^2}
[(1-\tau_2/\tau)(1-\tau_1/\tau)-(1-\tau_2/\tau)(\tau_2/\tau)+2(1-2\tau_2
/\tau)]\nonumber \\ \;\;\;\;
+\;O.T.
 \label{3.8}
\end{eqnarray}
A straightforward evaluation yields
\begin{equation}\label{3.9}
i\Delta^{(1)}\Gamma_{\rho}(p,q)
=(\frac{e}{2m}\sigma_{\rho\mu}(p-q)^{\mu}
)\frac{\alpha}{4\pi}\frac{m^2}{\Lambda^2}[-6.2986]\; +\; O.T.
\end{equation}
In a similar manner,we evaluate the diagrams ~\ref{sh-weak}(a),~\ref{sh-weak}(b)
,~\ref{sh-weak}(c).
We give the results.For the diagram of Fig.~\ref{sh-weak}(a),we obtain
\begin{eqnarray}
\lefteqn{i\Delta^{(2)}\Gamma_{\rho}(p,q)=(\frac{e}{2m}\sigma_{\rho\mu}
(p-q)^{\mu})\frac{e^2}{16\pi^2}\frac{m^2}{\Lambda^2}}
\nonumber \\ &\times&\left[\frac{-1}{\sin^2\theta_w}\right]
\int_0^1d\tau_1\int_0^1d\tau_2\int_0^1d\tau_3 \frac{1}{\tau^2}[2+
\frac{4\tau_1\tau_2}{\tau^2}
-\frac{12\tau_1}{\tau}]\; +\; O.T.
\label{3.10}
\end{eqnarray}

On simplification

\begin{eqnarray}
\lefteqn{i\Delta^{(2)}\Gamma_{\rho}(p,q)=}\nonumber \\ &\;&[\frac{e}{2m}
\sigma_{\rho\mu}(p-q)^{\mu}]
\frac{\alpha}{4\pi}\frac{m^2}{\Lambda^2}
\times[\frac{-1.4232}{\sin^2\theta_w}]\; +\; O.T.
\label{3.11}
\end{eqnarray}
For the diagrams of Figs.~\ref{sh-weak}(b) and ~\ref{sh-weak}(c) together,
we find
\begin{eqnarray}
\lefteqn{i\Delta^{(3)}\Gamma_{\rho}(p,q)=}\nonumber \\&\; &[\frac{e}{2m}
\sigma_{\rho\mu}(p-q)^{\mu}]
\frac{e^2}{16\pi^2}\frac{m^2}{\Lambda^2}\nonumber \\ &\times&[-\frac{1}
{\sin^2\theta_w}]
\int_0^1d\tau_1\int_0^1d\tau_2\int_0^1d\tau_3 \frac{\tau_2}
{\tau^3}\;+\;O.T.\nonumber \\ &=&
(\frac{e}{2m}\sigma_{\rho\mu}(p-q)^{\mu})
\frac{\alpha}{4\pi}\frac{m^2}{\Lambda^2}\left[\frac{-0.287}{\sin^2\theta_w}
\right]\; +\; O.T.
\label{3.12}
\end{eqnarray}
on simplification.

The diagram of Fig.~\ref{sh-weak}(d) clearly involves exactly 
one $\gamma$-matrix
and hence does not contribute to the anomalous magnetic moment.

The Z-exchange diagram of Fig.~\ref{sh-weak}(e) yields a contribution
\begin{eqnarray}
\lefteqn{i\Delta^{(4)}\Gamma_{\rho}(p,q)=(\frac{e}{2m}
\sigma_{\rho\mu}(p-q)^{\mu})
\frac{e^2}{16\pi^2}\frac{m^2}{\Lambda^2}
\frac{1}{\sin^2\theta_w \cos^4\theta_w}
\times\frac{1}{2}}\nonumber \\ &\times& [ (-1+4\sin^2\theta_w)^2
\times\int d\tau_1d\tau_2d\tau_3
 \frac{1}{\tau^2}(1-\frac{\tau_2}{\tau})(1-\frac{\tau_1}{\tau}-\frac{\tau_2}
 {\tau})\nonumber \\ & &+((-1+4\sin^2\theta_w)^2-1)
 \int d\tau_1 d\tau_2 d\tau_3\frac{1}{\tau^2}
  2(1-\frac{2\tau_2}{\tau})]
  \label{3.13} \\ &=& [
\frac{e}{2m}\sigma_{\rho\mu   
}(p-q)^{\mu}
\frac{e^2}{16\pi^2}\frac{m^2}{\Lambda^2}
 ] \times \frac{1}{\sin^2\theta_w \cos^4\theta_w}
 \times \nonumber \\ &\;&[0.7876(8\sin^4\theta_w -4\sin^2\theta_w)+0.1060]
 \; +\; O.T.
\label{3.14} \\ 
 &=&(\frac{e}{2m}\sigma_{\rho\mu}(p-q)^{\mu}
)\frac{\alpha}{4\pi}\frac{m^2}{\Lambda^2}[-2.0873]\; +\; O.T.
\label{3.15}
\end{eqnarray}
Finally we note that the diagrams of Figs.~\ref{sh-weak}(f)
and ~\ref{sh-weak}(g) are
suppresed by factors of $m^2/m_W^2$.These diagrams contribute 
finitely to the anomalous magnetic moment and {\em both} the
local and nonlocal contributions are suppressed by this additional 
factor;as an inspection of the contribution will show.We neglect 
them for present considerations.

The total shadow loop contribution for $a_\mu$ is given by the sum
of contributions from Eqs.\ref{3.9},\ref{3.11},\ref{3.12},\ref{3.15}
as
\begin{eqnarray}
 &=&(\frac{e}{2m}\sigma_{\rho\mu}(p-q)^{\mu}
)\frac{\alpha}{4\pi}\frac{m^2}{\Lambda^2}[-10.0961]\nonumber \\ &=&(
\frac{e}{2m}\sigma_{\rho\mu}(q-p)^{\mu}
)\frac{65.45(Mev)^2}{\Lambda^2}
\label{3.16}
\end{eqnarray}
From this we can read off the nonlocal electroweak corrections to 
$a_{\mu}$ of order $m^2/\Lambda^2$
\begin{equation} \label{3.17}
(\Delta a_{\mu})^{nl}_{th}\;  =\; \frac{(65.45)(Mev)^2}{\Lambda^2}
\end{equation}
The total theoretical contributions to $a_\mu$ from local theory
is\cite{fp}
\begin{equation}
a_{\mu}^{th}\; =\; 1165918  (2)\times 10^{-9}
\label{3.18}
\end{equation}
The experimental value of $a_{\mu}$ is presently\cite{fp}
\begin{equation} \label{3.19}
a_{\mu}^{exp}\; =\; 1165923 (8.5) \times 10^{-9}
\end{equation}
Therefore the contribution which may be attributed to nonlocal
corrections ,$\Delta$,is
\begin{equation}
\label{3.20}
\Delta=5\pm8.5\times 10^{-9}
\end{equation}
 We note that the nonlocal contribution to the anomalous magnetic moment 
 is positive and such as to close the gap between the local theoretical 
 value and the experimental one(if the numbers are taken literally,
 ignoring error bars).But in view of the fact that the scale of nonlocality
 ($1/\Lambda$),if it exists,may be quite small,this contribution is 
 not quite large
 enough to explain the (literal)difference.We further note that even for  a
 quite small $1/\Lambda$(of the order of $(500\;Gev)^{-1}$),the nonlocal
 contribution is comparable to the weak contribution.However,in view of the 
 error bars in \ref{3.20} we are unable at present,to obtain a stringent
 enough bound on $\Lambda$:We obtain
 \begin{equation}
 \label{3.21}
 \frac{1}{\Lambda} \; \leq\;3\times 10^{-16}.
 \end{equation}
 A similar bound has been obtained from nonlocal
 formulations using stochastic quantization.

 We in fact expect the bound to improve substantially once the 
 new experiment planned at Brookhaven National Laboratory ~\cite{vwh}
  to determine  $a_{\mu}$ is performed.The experimental error 
  is expected to come down  to $\pm 40\times  10^{-11}$,i.e,
  by a factor of 20 compared to the presently available data .
  Once the results of this experiment are available ,and the 
  hadronic contribution is calculated with greater accuracy \cite{holl}
  the bound of Eq.\ref{3.21} could be improved by a factor of upto
  5-6.

 We should however point out that the result \ref{3.17} has been 
 obtained on certain assumptions made in the calculations.Thus,the
 numerical result of \ref{3.17}is valid only if the actual scale
 of nonlocality $\Lambda^2 \gg M_W^2$;say $\Lambda>300\;Gev$.If
 $\Lambda$ were smaller than this ,corrections of the orders 
 $M_W^2/\Lambda^2$, $M_Z^2/\Lambda^2$ would become significant.
 We have not calculated these.However, we do expect these to 
 alter the result \ref{3.17} drastically so that the bound of \ref{3.21}
may still survive(modulo a factor not far from unity).
Of course once the results of the upcoming experiment mentioned 
above are available, the bound on $\Lambda$ is expected to be 
raised so that our approximations can be sustained.
 
\section{Comparison with nonstandard contributions}
\label{4}

As experimental accuracy with which (g-2) of the muon is measured
is improved in the latest experiments,comparison of these results
with the results expected from (local) Standard Model (calculated 
with improved accuracy), is expected to reveal much new physics.
Thus a discrepancy between these can be a signal of,say,nonlocality
of the underlying Standard Model as considered in this work or a signal of
new physics in addition to the Standard Model,viz. of the nonstandard effects
of various kinds.In this section we shall compare our results with 
several other works on additional contributions to (g-2) of the muon
due to these nonstandard effects~\cite{mery,Car,arzt}.

We shall compare our work successively with these works for their nonstandard 
physics considered in them and/or their methodology.While comparing these 
results we first make some general remarks.

(i)At the outset we point out a major difference:
Our work uses Standard Model fields with Standard Model interactions and adds
no other particles, and no   arbitrary  anomalous interactions 
to the Lagrangian except to nonlocalize the SM Lagrangian.
And this nonlocalization is carried out in a way restricted 
by the preservation of a (nonlocal) BRS invariance, 
renormalizability and unitarity of the theory(and is thus mostly free from
adhoc features.).In most of the works on nonstandard effects,on the other hand,
particles and/or  couplings, foreign to Standard Model are introduced.
This limits the scope of their comparison.

(ii)The second point one can make before going into 
specific works is as follows:
In many of the nonstandard effects, a new (higher)mass  scale $\Lambda_{NS}$
is involved.(This could be the scale of compositeness ,mass of additional new 
particles ,scale introduced in form factors etc.)In most cases, the additional 
contribution to (g-2) is of ${\cal O}(m^2_{\mu}/\Lambda^2_{NS})$.
If, then,the coupling 
involved is also of ${\cal O}(e)$,then, we would expect a bound 
on $\Lambda_{NS}$ of
the same order as on the scale of nonlocality $\Lambda$ in our work.

(iii)Finally we make a remark on our methodology.Calculations done 
in our work involve a renormalizable gauge always,and are done without 
adhoc cutoff and renormalization procedures.In the works on nonstandard 
effects~\cite{mery,arzt} ,on the other hand ,a necessity for 
introducing adhoc cutoff on momentum 
integrals(however physical)and an adhoc subtraction procedure (however
natural) arises.

With these remarks we proceed with specific comparisons.

 The work of Mery et al.~\cite{mery} takes into account 
 nonstandard effects of various
 kinds that can arize at lower energies from a (local) composite model with 
 a higher  scale of compositeness .Compositeness of leptons and gauge bosons 
 could lead to form factors for these.In addition,there could be excited 
 states of the Standard Model particles ($\mu^*,W^*,Z^* $ etc).There 
 could be residual effective 4-fermion interactions below the scale of 
 compositeness and nonstandard vector  boson couplings.
 The effect of form factors
 is taken into account by introducing a phenomenological form factor
 $(1-k^2/\Lambda^2_F)^{-1}$ and performing the calculations in Unitary 
 Gauge.The divergence of the integrals necessiates a physical cutoff
 $\Lambda \sim \Lambda_F$.The remarks (iii) and (ii) made in the 
 preceeding paragraph directly apply here.
 Comparison with results for diagrams involving excited 
 particles is difficult on account of remark (i) above except that remark 
 (ii)directly applies here.As for the residual 4-fermion interactions, 
 remarks (ii) and (iii) apply directly.It is, however,hard
 to compare the results
 for anomalous couplings of vector bosons as they have no analogue in our 
 work.

The work of Carena et al.~\cite{Car} discusses the corrections to (g-2)
in supersymmetric models.Here the corrections to (g-2) arise from 
additional diagrams involving supersymmetric partners of various 
standard model particles, and depend on their masses.This work then explores
bound on these masses and couplings.In this sense this work is completely different from ours in that we accept standard model as essentially correct but
for an allowed length scale $1/\Lambda$ and expect corrections from this scale.
On account of the entirely different  origins of the possible 
corrections to (g-2) a direct comparison of the results seem difficult;
except that one expects the scale $\Lambda$ in nonlocal theories and 
$\tilde{m}$,the scale of the masses of supersymmetric partners to be comparable.
(Note remark (i) and (ii) made earlier.)

The work of Arzt et al.~\cite{arzt} explores the corrections to (g-2) 
in a model 
independent way by formulating the 'non-Standard Model' terms as a series 
of dimension six operators in an Effective Lagrangian approach.It also 
naturally involves a scale $\Lambda$ at which these nonstandard corrections 
become significant.The calculation does require a cutoff procedure and 
additional renormalizations(involving extra renormalization conditions).
As per remark (ii) we would expect the scale $\Lambda$ 
involved in the effective
action to be comparable to our scale of nonlocality $\Lambda$ as the couplings
have been assumed to be of the same order as $e$.We however note several 
differences in methodology.

It may appear at first sight that the nonlocal W-S Lagrangian expanded to 
${\cal O}(1/\Lambda^2)$ is actually a special case of the effective Lagrangian 
approach albeit with known operators and known coefficients.In this connection
we point out two things.Firstly, if only ${\cal O}(1/\Lambda^2)$ terms were 
retained in our ${\cal L}$,the convergence of integrals that is present in our 
approach with full ${\cal L}$,would be lost and a need for adhoc cutoff
and renormalization procedure would be necessiated as in ~\cite{arzt}.Please
note remark (iii) made earlier.Secondly,the operators of ${\cal O}(1/\Lambda^2)$
arising in such an expansion of the nonlocal W-S action, would not be gauge 
invariant  as our action is invariant under a {\em nonlocal} BRS transformation
(i.e transformations themselves contain terms of various orders in 
$1/\Lambda^2$.)Of course the total action is BRS invariant.In this sense,
the assumptions of Arzt et al. about the gauge invariant nature of dimension
six operators is not directly fulfilled in such an expansion.

\begin{figure}
\begin{center} 
\begin{picture}(25000,16000)


\drawline\fermion[\E\REG](1000,4000)[8000]
\global\advance\pmidx by -100
\global\advance\pmidy by -500
\put(\pmidx,\pmidy){$\rule{0.7mm}{3.5mm}$}
\put(15000,4000){=$\;\;-i\int_0^1\frac{d\tau}{\Lambda^2}
e^{\frac{\tau}{\Lambda^2}(p^2-\mu^2)}$}
 

\drawline\fermion[\E\REG](1000,8000)[8000]
\put(15000,8000){=$\;\;-i\int_1^\infty\frac{d\tau}{\Lambda^2}
e^{\frac{\tau}{\Lambda^2}(p^2-\mu^2)}$}
\end{picture}
\end{center}
\caption{unbarred and barred propagators in nonlocal  $\phi^4$ theory}
\label{phi4prop}
\end{figure}
\begin{figure}
\begin{center} 
\begin{picture}(9000,6500)


\drawline\fermion[\E\REG](1000,1000)[8000]
\global\advance\pmidy by 2000
\put(\pmidx,\pmidy){\circle{4000}}
\end{picture}
\end{center}
\caption{self energy diagram in nonlocal  $\phi^4$ theory}
\label{uloop}
\end{figure}
\begin{figure}
\begin{center} 
\begin{picture}(9000,6500)


\drawline\fermion[\E\REG](1000,1000)[8000]
\global\advance\pmidy by 2000
\put(\pmidx,\pmidy){\circle{4000}}
\global\advance\pmidx by -100
\global\advance\pmidy by 1300
\put(\pmidx,\pmidy){$\rule{0.7mm}{3.5mm}$}
\end{picture}
\end{center}
\caption{shadow loop self energy diagram in nonlocal $\phi^4$ theory}
\label{bloop}
\end{figure}
\begin{figure}
\begin{picture}(35000,60000)

\put(1000,4000){$W(barred)$}
\drawline\photon[\E\REG](11000,4000)[8]
\global\advance\photonfronty by 500
\global\advance\photonbacky by 500
\put(\photonfrontx,\photonfronty){$\mu$}
\put(\photonbackx,\photonbacky){$\nu$}
\global\advance\pmidy by -700
\global\advance\pmidx by -100
\put(\pmidx,\pmidy){$\rule{0.7mm}{3.5mm}$}
\put(25000,4000){$+i\int_0^1\frac{d\tau}{\Lambda^2}
e^{\frac{\tau}{\Lambda^2}(k^2-m_W^2)}g_{\mu\nu}$}
 
\put(1000,8000){$W(smeared)$}
\drawline\photon[\E\REG](11000,8000)[8]
\global\advance\photonfronty by 500
\global\advance\photonbacky by 500
\put(\photonfrontx,\photonfronty){$\mu$}
\put(\photonbackx,\photonbacky){$\nu$}
\put(25000,8000){$+i\int_1^\infty\frac{d\tau}{\Lambda^2}
e^{\frac{\tau}{\Lambda^2}(k^2-m_W^2)}g_{\mu\nu}$}

\put(1000,12000){$Z(barred)$}
\drawline\photon[\E\REG](11000,12000)[8]
\global\advance\photonfronty by 500
\global\advance\photonbacky by 500
\put(\photonfrontx,\photonfronty){$\mu$}
\put(\photonbackx,\photonbacky){$\nu$}
\global\advance\pmidy by -700
\global\advance\pmidx by -100
\put(\pmidx,\pmidy){$\rule{0.7mm}{3.5mm}$}
\put(25000,12000){$+i\int_0^1\frac{d\tau}{\Lambda^2}
e^{\frac{\tau}{\Lambda^2}(k^2-m_z^2)}g_{\mu\nu}$}
 
\put(1000,16000){$Z(smeared)$}
\drawline\photon[\E\REG](11000,16000)[8]
\global\advance\photonfronty by 500
\global\advance\photonbacky by 500
\put(\photonfrontx,\photonfronty){$\mu$}
\put(\photonbackx,\photonbacky){$\nu$}
\put(25000,16000){$+i\int_1^\infty\frac{d\tau}{\Lambda^2}
e^{\frac{\tau}{\Lambda^2}(k^2-m_z^2)}g_{\mu\nu}$}

\put(1000,20000){$\gamma(barred)$}
\drawline\photon[\E\REG](11000,20000)[8]
\global\advance\photonfronty by 500
\global\advance\photonbacky by 500
\put(\photonfrontx,\photonfronty){$\mu$}
\put(\photonbackx,\photonbacky){$\nu$}
\global\advance\pmidy by -750
\global\advance\pmidx by -100
\put(\pmidx,\pmidy){$\rule{0.7mm}{3.5mm}$}
\put(25000,20000){$+i\int_0^1\frac{d\tau}{\Lambda^2}
e^{\frac{\tau}{\Lambda^2}k^2}g_{\mu\nu}$}
 
\put(1000,24000){$\gamma(smeared)$}
\drawline\photon[\E\REG](11000,24000)[8]
\global\advance\photonfronty by 500
\global\advance\photonbacky by 500
\put(\photonfrontx,\photonfronty){$\mu$}
\put(\photonbackx,\photonbacky){$\nu$}
\put(25000,24000){$+i\int_1^\infty\frac{d\tau}{\Lambda^2}
e^{\frac{\tau}{\Lambda^2}k^2}g_{\mu\nu}$}
\put(1000,28000){$\phi(barred)$}
\drawline\scalar[\E\REG](11000,28000)[4]
\global\advance\pmidy by -500
\global\advance\pmidx by -100
\put(\pmidx,\pmidy){$\rule{0.7mm}{3.5mm}$}
\put(25000,28000){$-i\int_0^1\frac{d\tau}{\Lambda^2}
e^{\frac{\tau}{\Lambda^2}(k^2-m_W^2)}$}
 
\put(1000,32000){$\phi(smeared)$}
\drawline\scalar[\E\REG](11000,32000)[4]
\put(25000,32000){$-i\int_1^\infty\frac{d\tau}{\Lambda^2}
e^{\frac{\tau}{\Lambda^2}(k^2-m_W^2)}$}
\put(1000,36000){$\phi_2(barred)$}
\drawline\scalar[\E\REG](11000,36000)[4]
\global\advance\pmidy by -500
\global\advance\pmidx by -100
\put(\pmidx,\pmidy){$\rule{0.7mm}{3.5mm}$}
\put(25000,36000){$-i\int_0^1\frac{d\tau}{\Lambda^2}
e^{\frac{\tau}{\Lambda^2}(k^2-m_z^2)}$}
 
\put(1000,40000){$\phi_2(smeared)$}
\drawline\scalar[\E\REG](11000,40000)[4]
\put(25000,40000){$-i\int_1^\infty\frac{d\tau}{\Lambda^2}
e^{\frac{\tau}{\Lambda^2}(k^2-m_z^2)}$}
\put(1000,44000){$\phi_1(barred)$}
\drawline\scalar[\E\REG](11000,44000)[4]
\global\advance\pmidy by -500
\global\advance\pmidx by -100
\put(\pmidx,\pmidy){$\rule{0.7mm}{3.5mm}$}
\put(25000,44000){$-i\int_0^1\frac{d\tau}{\Lambda^2}
e^{\frac{\tau}{\Lambda^2}(k^2-2\mu^2)}$}
 
\put(1000,48000){$\phi_1(smeared)$}
\drawline\scalar[\E\REG](11000,48000)[4]
\put(25000,48000){$-i\int_1^\infty\frac{d\tau}{\Lambda^2}
e^{\frac{\tau}{\Lambda^2}(p^2-2\mu^2)}$}

\put(1000,52000){$fermion(barred)$}
\drawline\fermion[\E\REG](11000,52000)[8000]
\global\advance\pmidx by -100
\global\advance\pmidy by -500
\put(\pmidx,\pmidy){$\rule{0.7mm}{3.5mm}$}
\global\advance\pmidx by -2000
\global\advance\pmidy by +500
\drawarrow[\LDIR\ATTIP](\pmidx,\pmidy)
\put(25000,52000){$-i\int_0^1\frac{d\tau}{\Lambda^2}
e^{\frac{\tau}{\Lambda^2}(k^2-m^2)}(\not k+m)$}
 
\put(1000,56000){$fermion(smeared)$}
\drawline\fermion[\E\REG](11000,56000)[8000]
\drawarrow[\LDIR\ATTIP](\pmidx,\pmidy)
\put(25000,56000){$-i\int_1^\infty\frac{d\tau}{\Lambda^2}
e^{\frac{\tau}{\Lambda^2}(k^2-m^2)}(\not k +m)$}
\end{picture}
\caption{ }
\label{props}
\end{figure}
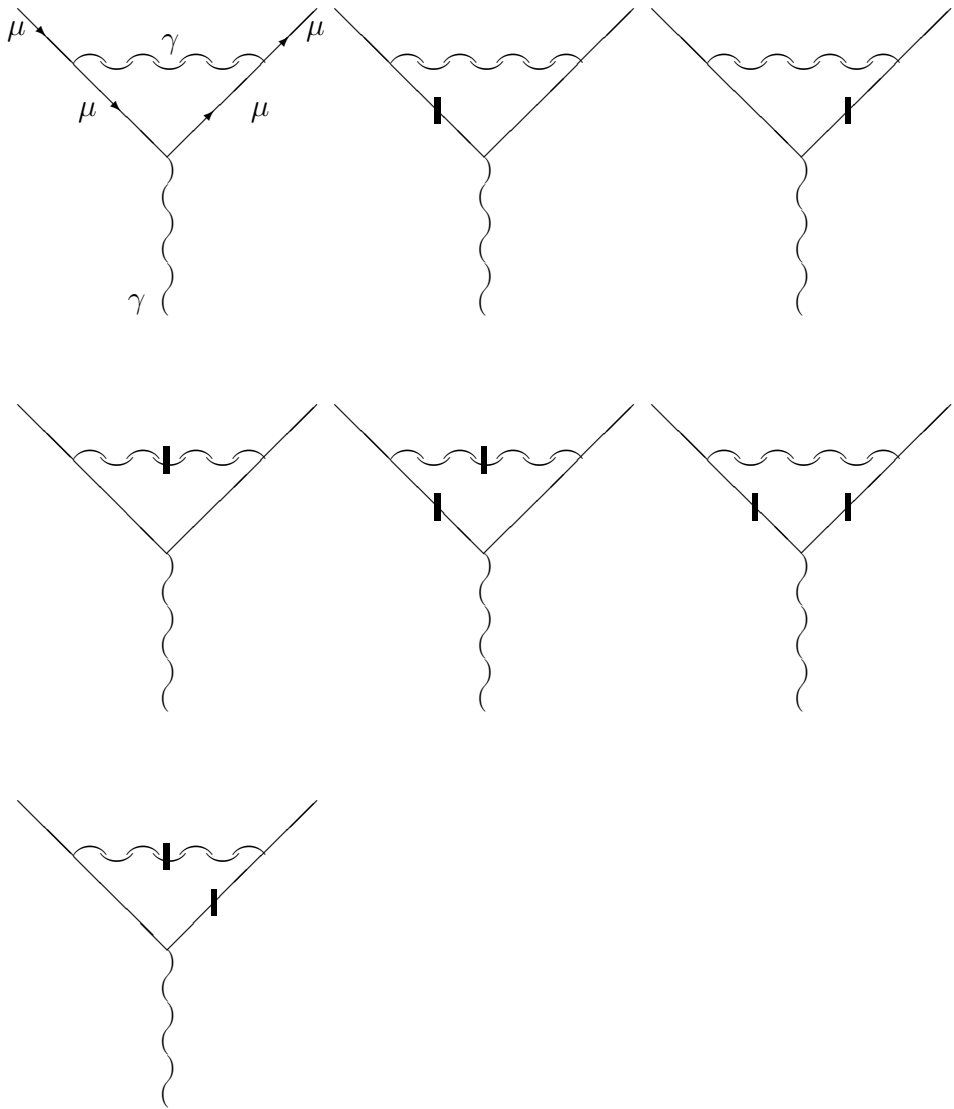
\begin{figure}
\begin{center}
\begin{picture}(36000,45000)
\drawline\fermion[\SE\REG](1000,42000)[3000]
\drawarrow[\LDIR\ATTIP](\pmidx,\pmidy)
\global\advance\fermionfrontx by -400
\global\advance\fermionfronty by -1000
\put(\fermionfrontx,\fermionfronty){$\mu$}
\drawline\photon[\E\REG](\fermionbackx,\fermionbacky)[7]
\global\advance\pmidy by 600
\global\advance\pmidx by -200
\put(\pmidx,\pmidy){$\gamma$}
\drawline\fermion[\SE\REG](\fermionbackx,\fermionbacky)[5000]
\drawarrow[\LDIR\ATTIP](\pmidx,\pmidy)
\global\advance\pmidx by -2000
\global\advance\pmidy by -200
\put(\pmidx,\pmidy){ $\mu$}
\drawline\photon[\S\REG](\fermionbackx,\fermionbacky)[6]
\global\advance\photonbackx by -1500
\global\advance\photonbacky by 300
\put(\photonbackx,\photonbacky){$\gamma$}
\drawline\fermion[\NE\REG](\fermionbackx,\fermionbacky)[5000]
\drawarrow[\LDIR\ATTIP](\pmidx,\pmidy)
\global\advance\pmidx by 1000
\global\advance\pmidy by -200
\put(\pmidx,\pmidy){ $\mu$}
\drawline\fermion[\NE\REG](\fermionbackx,\fermionbacky)[3000]
\drawarrow[\LDIR\ATTIP](\pmidx,\pmidy)
\global\advance\fermionbackx by -400
\global\advance\fermionbacky by -1000
\put(\fermionbackx,\fermionbacky){$\mu$}
\drawline\fermion[\SE\REG](13000,42000)[3000]
\drawline\photon[\E\REG](\fermionbackx,\fermionbacky)[7]
\drawline\fermion[\SE\REG](\fermionbackx,\fermionbacky)[5000]
\global\advance\pmidx by -100
\global\advance\pmidy by -500
\put(\pmidx,\pmidy){$\rule{.7mm}{3.5mm}$} 
\drawline\photon[\S\REG](\fermionbackx,\fermionbacky)[6]
\drawline\fermion[\NE\REG](\fermionbackx,\fermionbacky)[5000]
\drawline\fermion[\NE\REG](\fermionbackx,\fermionbacky)[3000]
\drawline\fermion[\SE\REG](25000,42000)[3000]
\drawline\photon[\E\REG](\fermionbackx,\fermionbacky)[7]
\drawline\fermion[\SE\REG](\fermionbackx,\fermionbacky)[5000]
\drawline\photon[\S\REG](\fermionbackx,\fermionbacky)[6]
\drawline\fermion[\NE\REG](\fermionbackx,\fermionbacky)[5000]
\global\advance\pmidx by -100
\global\advance\pmidy by -500
\put(\pmidx,\pmidy){$\rule{.7mm}{3.5mm}$}
\drawline\fermion[\NE\REG](\fermionbackx,\fermionbacky)[3000]
\drawline\fermion[\SE\REG](1000,27000)[3000]
\drawline\photon[\E\REG](\fermionbackx,\fermionbacky)[7]
\global\advance\pmidx by -100
\global\advance\pmidy by -500
\put(\pmidx,\pmidy){$\rule{.7mm}{3.5mm}$}
\drawline\fermion[\SE\REG](\fermionbackx,\fermionbacky)[5000]
\drawline\photon[\S\REG](\fermionbackx,\fermionbacky)[6]
\drawline\fermion[\NE\REG](\fermionbackx,\fermionbacky)[5000]
\drawline\fermion[\NE\REG](\fermionbackx,\fermionbacky)[3000]
\drawline\fermion[\SE\REG](13000,27000)[3000]
\drawline\photon[\E\REG](\fermionbackx,\fermionbacky)[7]
\global\advance\pmidx by -100
\global\advance\pmidy by -500
\put(\pmidx,\pmidy){$\rule{.7mm}{3.5mm}$}
\drawline\fermion[\SE\REG](\fermionbackx,\fermionbacky)[5000]
\global\advance\pmidx by -100
\global\advance\pmidy by -500
\put(\pmidx,\pmidy){$\rule{.7mm}{3.5mm}$}
\drawline\photon[\S\REG](\fermionbackx,\fermionbacky)[6]
\drawline\fermion[\NE\REG](\fermionbackx,\fermionbacky)[5000]
\drawline\fermion[\NE\REG](\fermionbackx,\fermionbacky)[3000]
\drawline\fermion[\SE\REG](25000,27000)[3000]
\drawline\photon[\E\REG](\fermionbackx,\fermionbacky)[7]
\drawline\fermion[\SE\REG](\fermionbackx,\fermionbacky)[5000]
\global\advance\pmidx by -100
\global\advance\pmidy by -500
\put(\pmidx,\pmidy){$\rule{.7mm}{3.5mm}$}
\drawline\photon[\S\REG](\fermionbackx,\fermionbacky)[6]
\drawline\fermion[\NE\REG](\fermionbackx,\fermionbacky)[5000]
\global\advance\pmidx by -100
\global\advance\pmidy by -500
\put(\pmidx,\pmidy){$\rule{.7mm}{3.5mm}$}
\drawline\fermion[\NE\REG](\fermionbackx,\fermionbacky)[3000]
\drawline\fermion[\SE\REG](1000,12000)[3000]
\drawline\photon[\E\REG](\fermionbackx,\fermionbacky)[7]
\global\advance\pmidx by -100
\global\advance\pmidy by -500
\put(\pmidx,\pmidy){$\rule{.7mm}{3.5mm}$}
\drawline\fermion[\SE\REG](\fermionbackx,\fermionbacky)[5000]
\drawline\photon[\S\REG](\fermionbackx,\fermionbacky)[6]
\drawline\fermion[\NE\REG](\fermionbackx,\fermionbacky)[5000]
\global\advance\pmidx by -100
\global\advance\pmidy by -500
\put(\pmidx,\pmidy){$\rule{.7mm}{3.5mm}$}
\drawline\fermion[\NE\REG](\fermionbackx,\fermionbacky)[3000]
\end{picture}
\end{center}
\caption{  }
\label{qedloops}
\end{figure}  
%
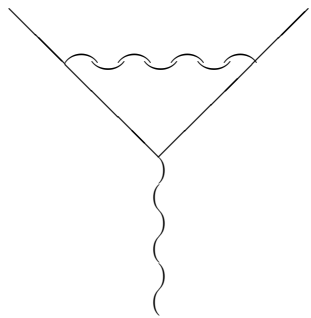
\begin{figure}
\begin{center}
\begin{picture}(30000,15000)
\drawline\fermion[\SE\REG](1000,12000)[3000]
\drawline\photon[\E\REG](\fermionbackx,\fermionbacky)[7]
\drawline\fermion[\SE\REG](\fermionbackx,\fermionbacky)[5000]
\drawline\photon[\S\REG](\fermionbackx,\fermionbacky)[6]
\drawline\fermion[\NE\REG](\fermionbackx,\fermionbacky)[5000]
\drawline\fermion[\NE\REG](\fermionbackx,\fermionbacky)[3000]
\put(15000,6000){$+\ \ barred\ \  variations$}
\end{picture}
\end{center}
\caption{  }
\label{fvert}
\end{figure}
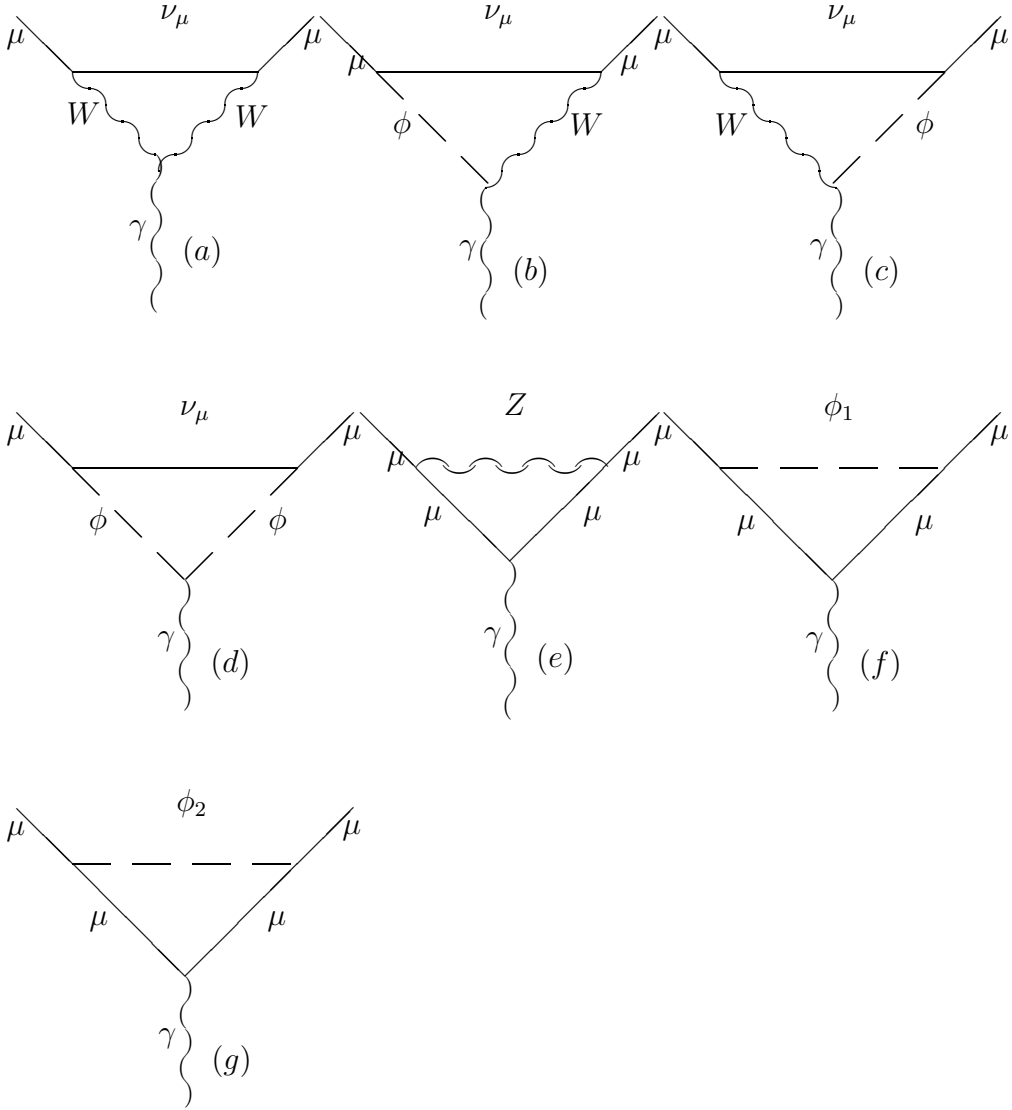
\begin{figure}
\begin{picture}(36000,45000)
\drawline\fermion[\SE\REG](500,42000)[3000]
\global\advance\fermionfrontx by -400
\global\advance\fermionfronty by -1000
\put(\fermionfrontx,\fermionfronty){$\mu$}
\drawline\photon[\SE\FLIPPED](\fermionbackx,\fermionbacky)[5]
\global\advance\pmidx by -1800
\global\advance\pmidy by -400
\put(\pmidx,\pmidy){$W$}
\drawline\fermion[\E\REG](\fermionbackx,\fermionbacky)[7000]
\global\advance\pmidy by 2000
\global\advance\pmidx by -200
\put(\pmidx,\pmidy){$\nu_{\mu}$}
\drawline\photon[\S\REG](\photonbackx,\photonbacky)[6]
\global\advance\pmidx by -1000
\put(\pmidx,\pmidy){$\gamma$}
\global\advance\pmidy by -1000
\global\advance\pmidx by 2000
\put(\pmidx,\pmidy){$(a)$}
\drawline\photon[\SW\REG](\fermionbackx,\fermionbacky)[6]
\global\advance\pmidx by 1000
\global\advance\pmidy by -200
\put(\pmidx,\pmidy){$W$}
\drawline\fermion[\NE\REG](\fermionbackx,\fermionbacky)[3000]
\global\advance\fermionbackx by -400
\global\advance\fermionbacky by -1000
\put(\fermionbackx,\fermionbacky){$\mu$}
\drawline\fermion[\SE\REG](12000,42000)[3000]
\global\advance\fermionfrontx by 1000
\global\advance\fermionfronty by -2000
\put(\fermionfrontx,\fermionfronty){$\mu$}
\drawline\scalar[\SE\REG](\fermionbackx,\fermionbacky)[3]
\global\advance\pmidx by -1500
\global\advance\pmidy by -200
\put(\pmidx,\pmidy){$\phi$}
\drawline\fermion[\E\REG](\fermionbackx,\fermionbacky)[8490]
\global\advance\pmidy by 2000
\global\advance\pmidx by -200
\put(\pmidx,\pmidy){$\nu_{\mu}$}
\drawline\photon[\SW\REG](\fermionbackx,\fermionbacky)[7]
\global\advance\pmidx by 1000
\global\advance\pmidy by -200
\put(\pmidx,\pmidy){$W$}
\drawline\photon[\S\FLIPPED](\photonbackx,\photonbacky)[5]
\global\advance\pmidx by -1000
\put(\pmidx,\pmidy){$\gamma$}
\global\advance\pmidy by -1000
\global\advance\pmidx by 2000
\put(\pmidx,\pmidy){$(b)$}
\drawline\fermion[\NE\REG](\fermionbackx,\fermionbacky)[3000]
\global\advance\fermionbackx by -1400
\global\advance\fermionbacky by -2000
\put(\fermionbackx,\fermionbacky){$\mu$}
\drawline\fermion[\SE\REG](25000,42000)[3000]
\global\advance\fermionfrontx by -400
\global\advance\fermionfronty by -1000
\put(\fermionfrontx,\fermionfronty){$\mu$}
\drawline\photon[\SE\FLIPPED](\fermionbackx,\fermionbacky)[7]
\global\advance\pmidx by -2400
\global\advance\pmidy by -200
\put(\pmidx,\pmidy){$W$}
\drawline\fermion[\E\REG](\fermionbackx,\fermionbacky)[8490]
\global\advance\pmidy by 2000
\global\advance\pmidx by -200
\put(\pmidx,\pmidy){$\nu_{\mu}$}
\drawline\photon[\S\REG](\photonbackx,\photonbacky)[5]
\global\advance\pmidx by -1000
\put(\pmidx,\pmidy){$\gamma$}
\global\advance\pmidy by -1000
\global\advance\pmidx by 2000
\put(\pmidx,\pmidy){$(c)$}
\drawline\scalar[\SW\REG](\fermionbackx,\fermionbacky)[3]
\global\advance\pmidx by 1000
\global\advance\pmidy by -200
\put(\pmidx,\pmidy){$\phi$}
\drawline\fermion[\NE\REG](\fermionbackx,\fermionbacky)[3000]
\global\advance\fermionbackx by -400
\global\advance\fermionbacky by -1000
\put(\fermionbackx,\fermionbacky){$\mu$}
\drawline\fermion[\SE\REG](500,27000)[3000]
\global\advance\fermionfrontx by -400
\global\advance\fermionfronty by -1000
\put(\fermionfrontx,\fermionfronty){$\mu$}
\drawline\scalar[\SE\REG](\fermionbackx,\fermionbacky)[3]
\global\advance\pmidx by -1500
\global\advance\pmidy by -200
\put(\pmidx,\pmidy){$\phi$}
\drawline\fermion[\E\REG](\fermionbackx,\fermionbacky)[8490]
\global\advance\pmidy by 2000
\global\advance\pmidx by -200
\put(\pmidx,\pmidy){$\nu_{\mu}$}
\drawline\photon[\S\REG](\scalarbackx,\scalarbacky)[5]
\global\advance\pmidx by -1000
\put(\pmidx,\pmidy){$\gamma$}
\global\advance\pmidy by -1000
\global\advance\pmidx by 2000
\put(\pmidx,\pmidy){$(d)$}
\drawline\scalar[\SW\REG](\fermionbackx,\fermionbacky)[3]
\global\advance\pmidx by 1000
\global\advance\pmidy by -200
\put(\pmidx,\pmidy){$\phi$}
\drawline\fermion[\NE\REG](\fermionbackx,\fermionbacky)[3000]
\global\advance\fermionbackx by -400
\global\advance\fermionbacky by -1000
\put(\fermionbackx,\fermionbacky){$\mu$}
\drawline\fermion[\SE\REG](13500,27000)[3000]
\global\advance\fermionfrontx by 1000
\global\advance\fermionfronty by -2000
\put(\fermionfrontx,\fermionfronty){$\mu$}
\drawline\photon[\E\REG](\fermionbackx,\fermionbacky)[7]
\global\advance\pmidy by 2000
\global\advance\pmidx by -200
\put(\pmidx,\pmidy){$Z$}
\drawline\fermion[\SE\REG](\fermionbackx,\fermionbacky)[5000]
\global\advance\pmidx by -1500
\global\advance\pmidy by -200
\put(\pmidx,\pmidy){$\mu$}
\drawline\photon[\S\REG](\fermionbackx,\fermionbacky)[6]
\global\advance\pmidx by -1000
\put(\pmidx,\pmidy){$\gamma$}
\global\advance\pmidy by -1000
\global\advance\pmidx by 2000
\put(\pmidx,\pmidy){$(e)$}
\drawline\fermion[\NE\REG](\fermionbackx,\fermionbacky)[5000]
\global\advance\pmidx by 1000
\global\advance\pmidy by -200
\put(\pmidx,\pmidy){$\mu$}
\drawline\fermion[\NE\REG](\fermionbackx,\fermionbacky)[3000]
\global\advance\fermionbackx by -1400
\global\advance\fermionbacky by -2000
\put(\fermionbackx,\fermionbacky){$\mu$} 
\drawline\fermion[\SE\REG](25000,27000)[3000]
\global\advance\fermionfrontx by -400
\global\advance\fermionfronty by -1000
\put(\fermionfrontx,\fermionfronty){$\mu$}
\drawline\scalar[\E\REG](\fermionbackx,\fermionbacky)[4]
\global\advance\pmidy by 2000
\global\advance\pmidx by -200
\put(\pmidx,\pmidy){$\phi_1$}
\drawline\fermion[\SE\REG](\fermionbackx,\fermionbacky)[6000]
\global\advance\pmidx by -1500
\global\advance\pmidy by -200
\put(\pmidx,\pmidy){$\mu$}
\drawline\photon[\S\REG](\fermionbackx,\fermionbacky)[5]
\global\advance\pmidx by -1000
\put(\pmidx,\pmidy){$\gamma$}
\global\advance\pmidy by -1000
\global\advance\pmidx by 2000
\put(\pmidx,\pmidy){$(f)$}
\drawline\fermion[\NE\REG](\fermionbackx,\fermionbacky)[6000]
\global\advance\pmidx by 1000
\global\advance\pmidy by -200
\put(\pmidx,\pmidy){$\mu$}
\drawline\fermion[\NE\REG](\fermionbackx,\fermionbacky)[3000]
\global\advance\fermionbackx by -400
\global\advance\fermionbacky by -1000
\put(\fermionbackx,\fermionbacky){$\mu$} 
\drawline\fermion[\SE\REG](500,12000)[3000]
\global\advance\fermionfrontx by -400
\global\advance\fermionfronty by -1000
\put(\fermionfrontx,\fermionfronty){$\mu$}
\drawline\scalar[\E\REG](\fermionbackx,\fermionbacky)[4]
\global\advance\pmidy by 2000
\global\advance\pmidx by -200
\put(\pmidx,\pmidy){$\phi_2$}
\drawline\fermion[\SE\REG](\fermionbackx,\fermionbacky)[6000]
\global\advance\pmidx by -1500
\global\advance\pmidy by -200
\put(\pmidx,\pmidy){$\mu$}
\drawline\photon[\S\REG](\fermionbackx,\fermionbacky)[5]
\global\advance\pmidx by -1000
\put(\pmidx,\pmidy){$\gamma$}
\global\advance\pmidy by -1000
\global\advance\pmidx by 2000
\put(\pmidx,\pmidy){$(g)$}
\drawline\fermion[\NE\REG](\fermionbackx,\fermionbacky)[6000]
\global\advance\pmidx by 1000
\global\advance\pmidy by -200
\put(\pmidx,\pmidy){$\mu$}
\drawline\fermion[\NE\REG](\fermionbackx,\fermionbacky)[3000]
\global\advance\fermionbackx by -400
\global\advance\fermionbacky by -1000
\put(\fermionbackx,\fermionbacky){$\mu$} 
\end{picture}
\caption{  }
\label{sm-weak}
\end{figure}  
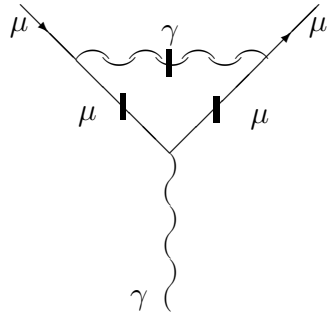
\begin{figure}
\begin{center}
\begin{picture}(10000,15000)
\drawline\fermion[\SE\REG](1000,12000)[3000]
\drawarrow[\LDIR\ATTIP](\pmidx,\pmidy)
\global\advance\fermionfrontx by -400
\global\advance\fermionfronty by -1000
\put(\fermionfrontx,\fermionfronty){$\mu$}
\drawline\photon[\E\REG](\fermionbackx,\fermionbacky)[7]
\global\advance\pmidx by -100
\global\advance\pmidy by -600
\put(\pmidx,\pmidy){$\rule{.7mm}{3.5mm}$} 
\global\advance\pmidy by 1400
\global\advance\pmidx by -200
\put(\pmidx,\pmidy){$\gamma$}
\drawline\fermion[\SE\REG](\fermionbackx,\fermionbacky)[5000]
\global\advance\pmidx by -100
\global\advance\pmidy by -500
\put(\pmidx,\pmidy){$\rule{.7mm}{3.5mm}$} 
\global\advance\pmidx by -2000
\put(\pmidx,\pmidy){ $\mu$}
\drawline\photon[\S\REG](\fermionbackx,\fermionbacky)[6]
\global\advance\photonbackx by -1500
\global\advance\photonbacky by 300
\put(\photonbackx,\photonbacky){$\gamma$}
\drawline\fermion[\NE\REG](\fermionbackx,\fermionbacky)[5000]
\global\advance\pmidx by -100
\global\advance\pmidy by -600
\put(\pmidx,\pmidy){$\rule{.7mm}{3.5mm}$} 
\global\advance\pmidx by 1000
\put(\pmidx,\pmidy){ $\mu$}
\drawline\fermion[\NE\REG](\fermionbackx,\fermionbacky)[3000]
\drawarrow[\LDIR\ATTIP](\pmidx,\pmidy)
\global\advance\fermionbackx by -400
\global\advance\fermionbacky by -1000
\put(\fermionbackx,\fermionbacky){$\mu$}
\end{picture}
\end{center}
\caption{  }
\label{sh.qed}
\end{figure}  
%
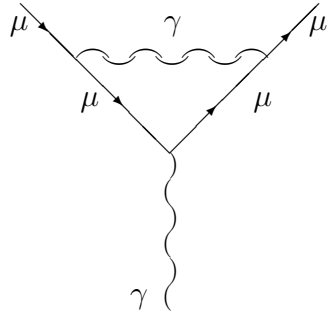
\begin{figure}
\begin{center}
\begin{picture}(10000,15000)
\drawline\fermion[\SE\REG](1000,12000)[3000]
\drawarrow[\LDIR\ATTIP](\pmidx,\pmidy)
\global\advance\fermionfrontx by -400
\global\advance\fermionfronty by -1000
\put(\fermionfrontx,\fermionfronty){$\mu$}
\drawline\photon[\E\REG](\fermionbackx,\fermionbacky)[7]
\global\advance\pmidy by 1100
\global\advance\pmidx by -200
\put(\pmidx,\pmidy){$\gamma$}
\drawline\fermion[\SE\REG](\fermionbackx,\fermionbacky)[5000]
\drawarrow[\LDIR\ATTIP](\pmidx,\pmidy)
\global\advance\pmidx by -2000
\put(\pmidx,\pmidy){ $\mu$}
\drawline\photon[\S\REG](\fermionbackx,\fermionbacky)[6]
\global\advance\photonbackx by -1500
\global\advance\photonbacky by 300
\put(\photonbackx,\photonbacky){$\gamma$}
\drawline\fermion[\NE\REG](\fermionbackx,\fermionbacky)[5000]
\drawarrow[\LDIR\ATTIP](\pmidx,\pmidy)
\global\advance\pmidx by 1000
\put(\pmidx,\pmidy){ $\mu$}
\drawline\fermion[\NE\REG](\fermionbackx,\fermionbacky)[3000]
\drawarrow[\LDIR\ATTIP](\pmidx,\pmidy)
\global\advance\fermionbackx by -400
\global\advance\fermionbacky by -1000
\put(\fermionbackx,\fermionbacky){$\mu$}
\end{picture}
\end{center}
\caption{  }
\label{l.qed}
\end{figure}  
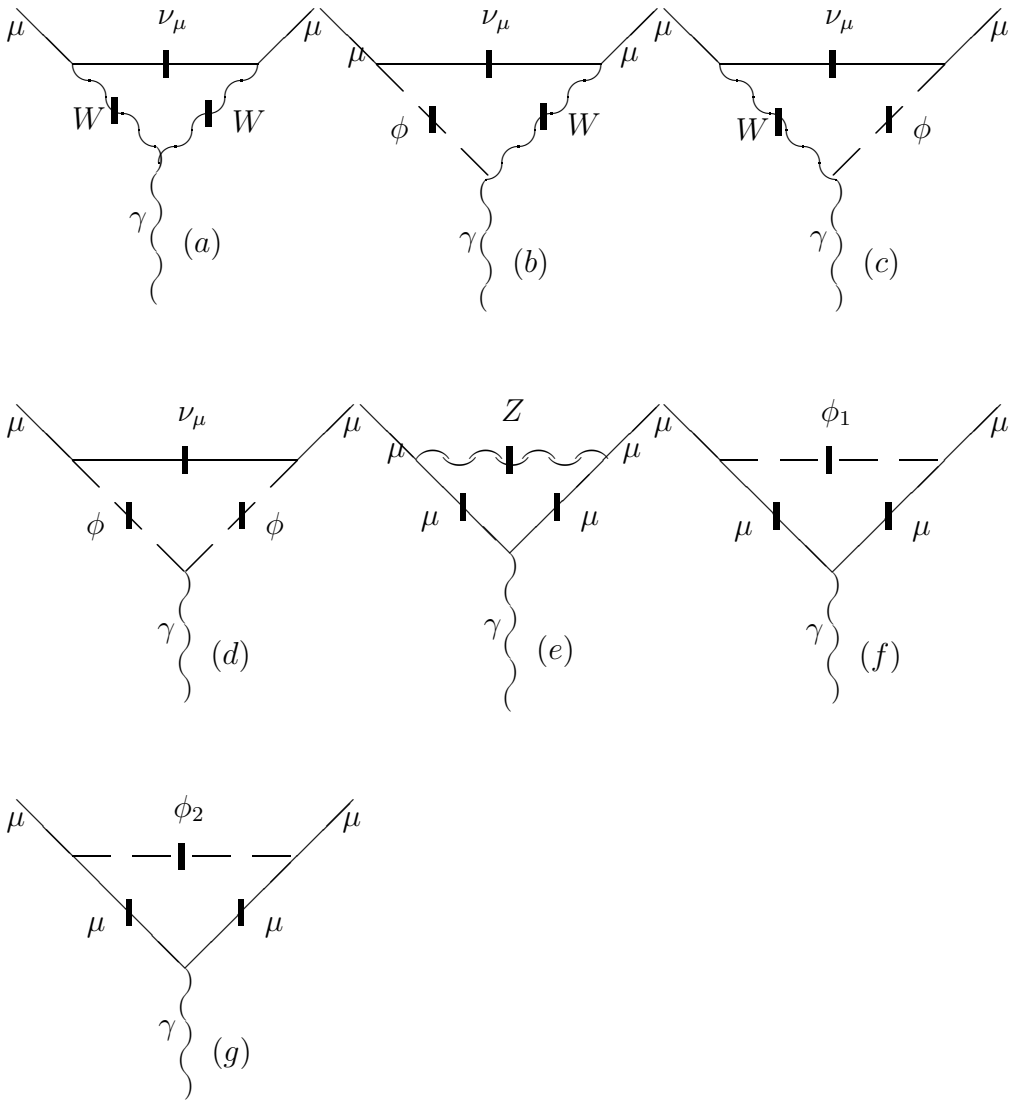
\begin{figure}
\begin{picture}(36000,45000)
\drawline\fermion[\SE\REG](500,42000)[3000]
\global\advance\fermionfrontx by -400
\global\advance\fermionfronty by -1000
\put(\fermionfrontx,\fermionfronty){$\mu$}
\drawline\photon[\SE\FLIPPED](\fermionbackx,\fermionbacky)[5]
\global\advance\pmidx by -100
\global\advance\pmidy by -700
\put(\pmidx,\pmidy){$\rule{.7mm}{3.5mm}$} 
\global\advance\pmidx by -1500
\global\advance\pmidy by -200
\put(\pmidx,\pmidy){$W$}
\drawline\fermion[\E\REG](\fermionbackx,\fermionbacky)[7000]
\global\advance\pmidx by -100
\global\advance\pmidy by -500
\put(\pmidx,\pmidy){$\rule{.7mm}{3.5mm}$} 
\global\advance\pmidy by 2000
\global\advance\pmidx by -200
\put(\pmidx,\pmidy){$\nu_{\mu}$}
\drawline\photon[\S\REG](\photonbackx,\photonbacky)[6]
\global\advance\pmidx by -1000
\put(\pmidx,\pmidy){$\gamma$}
\global\advance\pmidy by -1000
\global\advance\pmidx by 2000
\put(\pmidx,\pmidy){$(a)$}
\drawline\photon[\SW\REG](\fermionbackx,\fermionbacky)[6]
\global\advance\pmidx by -100
\global\advance\pmidy by -500
\put(\pmidx,\pmidy){$\rule{.7mm}{3.5mm}$} 
\global\advance\pmidx by 1000
\global\advance\pmidy by -200
\put(\pmidx,\pmidy){$W$}
\drawline\fermion[\NE\REG](\fermionbackx,\fermionbacky)[3000]
\global\advance\fermionbackx by -400
\global\advance\fermionbacky by -1000
\put(\fermionbackx,\fermionbacky){$\mu$}
\drawline\fermion[\SE\REG](12000,42000)[3000]
\global\advance\fermionfrontx by 1000
\global\advance\fermionfronty by -2000
\put(\fermionfrontx,\fermionfronty){$\mu$}
\drawline\scalar[\SE\REG](\fermionbackx,\fermionbacky)[3]
\global\advance\pmidx by -100
\global\advance\pmidy by -500
\put(\pmidx,\pmidy){$\rule{.7mm}{3.5mm}$} 
\global\advance\pmidx by -1500
\global\advance\pmidy by -200
\put(\pmidx,\pmidy){$\phi$}
\drawline\fermion[\E\REG](\fermionbackx,\fermionbacky)[8490]
\global\advance\pmidx by -100
\global\advance\pmidy by -500
\put(\pmidx,\pmidy){$\rule{.7mm}{3.5mm}$} 
\global\advance\pmidy by 2000
\global\advance\pmidx by -200
\put(\pmidx,\pmidy){$\nu_{\mu}$}
\drawline\photon[\SW\REG](\fermionbackx,\fermionbacky)[7]
\global\advance\pmidx by -100
\global\advance\pmidy by -300
\put(\pmidx,\pmidy){$\rule{.7mm}{3.5mm}$} 
\global\advance\pmidx by 1000
\global\advance\pmidy by -200
\put(\pmidx,\pmidy){$W$}
\drawline\photon[\S\FLIPPED](\photonbackx,\photonbacky)[5]
\global\advance\pmidx by -1000
\put(\pmidx,\pmidy){$\gamma$}
\global\advance\pmidy by -1000
\global\advance\pmidx by 2000
\put(\pmidx,\pmidy){$(b)$}
\drawline\fermion[\NE\REG](\fermionbackx,\fermionbacky)[3000]
\global\advance\fermionbackx by -1400
\global\advance\fermionbacky by -2000
\put(\fermionbackx,\fermionbacky){$\mu$}
\drawline\fermion[\SE\REG](25000,42000)[3000]
\global\advance\fermionfrontx by -400
\global\advance\fermionfronty by -1000
\put(\fermionfrontx,\fermionfronty){$\mu$}
\drawline\photon[\SE\FLIPPED](\fermionbackx,\fermionbacky)[7]
\global\advance\pmidx by -100
\global\advance\pmidy by -500
\put(\pmidx,\pmidy){$\rule{.7mm}{3.5mm}$} 
\global\advance\pmidx by -1500
\global\advance\pmidy by -200
\put(\pmidx,\pmidy){$W$}
\drawline\fermion[\E\REG](\fermionbackx,\fermionbacky)[8490]
\global\advance\pmidx by -100
\global\advance\pmidy by -500
\put(\pmidx,\pmidy){$\rule{.7mm}{3.5mm}$} 
\global\advance\pmidy by 2000
\global\advance\pmidx by -200
\put(\pmidx,\pmidy){$\nu_{\mu}$}
\drawline\photon[\S\REG](\photonbackx,\photonbacky)[5]
\global\advance\pmidx by -1000
\put(\pmidx,\pmidy){$\gamma$}
\global\advance\pmidy by -1000
\global\advance\pmidx by 2000
\put(\pmidx,\pmidy){$(c)$}
\drawline\scalar[\SW\REG](\fermionbackx,\fermionbacky)[3]
\global\advance\pmidx by -100
\global\advance\pmidy by -500
\put(\pmidx,\pmidy){$\rule{.7mm}{3.5mm}$} 
\global\advance\pmidx by 1000
\global\advance\pmidy by -200
\put(\pmidx,\pmidy){$\phi$}
\drawline\fermion[\NE\REG](\fermionbackx,\fermionbacky)[3000]
\global\advance\fermionbackx by -400
\global\advance\fermionbacky by -1000
\put(\fermionbackx,\fermionbacky){$\mu$}
\drawline\fermion[\SE\REG](500,27000)[3000]
\global\advance\fermionfrontx by -400
\global\advance\fermionfronty by -1000
\put(\fermionfrontx,\fermionfronty){$\mu$}
\drawline\scalar[\SE\REG](\fermionbackx,\fermionbacky)[3]
\global\advance\pmidx by -100
\global\advance\pmidy by -500
\put(\pmidx,\pmidy){$\rule{.7mm}{3.5mm}$} 
\global\advance\pmidx by -1500
\global\advance\pmidy by -200
\put(\pmidx,\pmidy){$\phi$}
\drawline\fermion[\E\REG](\fermionbackx,\fermionbacky)[8490]
\global\advance\pmidx by -100
\global\advance\pmidy by -500
\put(\pmidx,\pmidy){$\rule{.7mm}{3.5mm}$} 
\global\advance\pmidy by 2000
\global\advance\pmidx by -200
\put(\pmidx,\pmidy){$\nu_{\mu}$}
\drawline\photon[\S\REG](\scalarbackx,\scalarbacky)[5]
\global\advance\pmidx by -1000
\put(\pmidx,\pmidy){$\gamma$}
\global\advance\pmidy by -1000
\global\advance\pmidx by 2000
\put(\pmidx,\pmidy){$(d)$}
\drawline\scalar[\SW\REG](\fermionbackx,\fermionbacky)[3]
\global\advance\pmidx by -100
\global\advance\pmidy by -500
\put(\pmidx,\pmidy){$\rule{.7mm}{3.5mm}$} 
\global\advance\pmidx by 1000
\global\advance\pmidy by -200
\put(\pmidx,\pmidy){$\phi$}
\drawline\fermion[\NE\REG](\fermionbackx,\fermionbacky)[3000]
\global\advance\fermionbackx by -400
\global\advance\fermionbacky by -1000
\put(\fermionbackx,\fermionbacky){$\mu$}
\drawline\fermion[\SE\REG](13500,27000)[3000]
\global\advance\fermionfrontx by 1000
\global\advance\fermionfronty by -2000
\put(\fermionfrontx,\fermionfronty){$\mu$}
\drawline\photon[\E\REG](\fermionbackx,\fermionbacky)[7]
\global\advance\pmidx by -100
\global\advance\pmidy by -500
\put(\pmidx,\pmidy){$\rule{.7mm}{3.5mm}$} 
\global\advance\pmidy by 2000
\global\advance\pmidx by -200
\put(\pmidx,\pmidy){$Z$}
\drawline\fermion[\SE\REG](\fermionbackx,\fermionbacky)[5000]
\global\advance\pmidx by -100
\global\advance\pmidy by -500
\put(\pmidx,\pmidy){$\rule{.7mm}{3.5mm}$} 
\global\advance\pmidx by -1500
\global\advance\pmidy by -200
\put(\pmidx,\pmidy){$\mu$}
\drawline\photon[\S\REG](\fermionbackx,\fermionbacky)[6]
\global\advance\pmidx by -1000
\put(\pmidx,\pmidy){$\gamma$}
\global\advance\pmidy by -1000
\global\advance\pmidx by 2000
\put(\pmidx,\pmidy){$(e)$}
\drawline\fermion[\NE\REG](\fermionbackx,\fermionbacky)[5000]
\global\advance\pmidx by -100
\global\advance\pmidy by -500
\put(\pmidx,\pmidy){$\rule{.7mm}{3.5mm}$} 
\global\advance\pmidx by 1000
\global\advance\pmidy by -200
\put(\pmidx,\pmidy){$\mu$}
\drawline\fermion[\NE\REG](\fermionbackx,\fermionbacky)[3000]
\global\advance\fermionbackx by -1400
\global\advance\fermionbacky by -2000
\put(\fermionbackx,\fermionbacky){$\mu$} 
\drawline\fermion[\SE\REG](25000,27000)[3000]
\global\advance\fermionfrontx by -400
\global\advance\fermionfronty by -1000
\put(\fermionfrontx,\fermionfronty){$\mu$}
\drawline\scalar[\E\REG](\fermionbackx,\fermionbacky)[4]
\global\advance\pmidx by -100
\global\advance\pmidy by -500
\put(\pmidx,\pmidy){$\rule{.7mm}{3.5mm}$} 
\global\advance\pmidy by 2000
\global\advance\pmidx by -200
\put(\pmidx,\pmidy){$\phi_1$}
\drawline\fermion[\SE\REG](\fermionbackx,\fermionbacky)[6000]
\global\advance\pmidx by -100
\global\advance\pmidy by -500
\put(\pmidx,\pmidy){$\rule{.7mm}{3.5mm}$} 
\global\advance\pmidx by -1500
\global\advance\pmidy by -200
\put(\pmidx,\pmidy){$\mu$}
\drawline\photon[\S\REG](\fermionbackx,\fermionbacky)[5]
\global\advance\pmidx by -1000
\put(\pmidx,\pmidy){$\gamma$}
\global\advance\pmidy by -1000
\global\advance\pmidx by 2000
\put(\pmidx,\pmidy){$(f)$}
\drawline\fermion[\NE\REG](\fermionbackx,\fermionbacky)[6000]
\global\advance\pmidx by -100
\global\advance\pmidy by -500
\put(\pmidx,\pmidy){$\rule{.7mm}{3.5mm}$} 
\global\advance\pmidx by 1000
\global\advance\pmidy by -200
\put(\pmidx,\pmidy){$\mu$}
\drawline\fermion[\NE\REG](\fermionbackx,\fermionbacky)[3000]
\global\advance\fermionbackx by -400
\global\advance\fermionbacky by -1000
\put(\fermionbackx,\fermionbacky){$\mu$} 
\drawline\fermion[\SE\REG](500,12000)[3000]
\global\advance\fermionfrontx by -400
\global\advance\fermionfronty by -1000
\put(\fermionfrontx,\fermionfronty){$\mu$}
\drawline\scalar[\E\REG](\fermionbackx,\fermionbacky)[4]
\global\advance\pmidx by -100
\global\advance\pmidy by -500
\put(\pmidx,\pmidy){$\rule{.7mm}{3.5mm}$} 
\global\advance\pmidy by 2000
\global\advance\pmidx by -200
\put(\pmidx,\pmidy){$\phi_2$}
\drawline\fermion[\SE\REG](\fermionbackx,\fermionbacky)[6000]
\global\advance\pmidx by -100
\global\advance\pmidy by -500
\put(\pmidx,\pmidy){$\rule{.7mm}{3.5mm}$} 
\global\advance\pmidx by -1500
\global\advance\pmidy by -200
\put(\pmidx,\pmidy){$\mu$}
\drawline\photon[\S\REG](\fermionbackx,\fermionbacky)[5]
\global\advance\pmidx by -1000
\put(\pmidx,\pmidy){$\gamma$}
\global\advance\pmidy by -1000
\global\advance\pmidx by 2000
\put(\pmidx,\pmidy){$(g)$}
\drawline\fermion[\NE\REG](\fermionbackx,\fermionbacky)[6000]
\global\advance\pmidx by -100
\global\advance\pmidy by -500
\put(\pmidx,\pmidy){$\rule{.7mm}{3.5mm}$} 
\global\advance\pmidx by 1000
\global\advance\pmidy by -200
\put(\pmidx,\pmidy){$\mu$}
\drawline\fermion[\NE\REG](\fermionbackx,\fermionbacky)[3000]
\global\advance\fermionbackx by -400
\global\advance\fermionbacky by -1000
\put(\fermionbackx,\fermionbacky){$\mu$} 
\end{picture}
\caption{  }
\label{sh-weak}
\end{figure}  
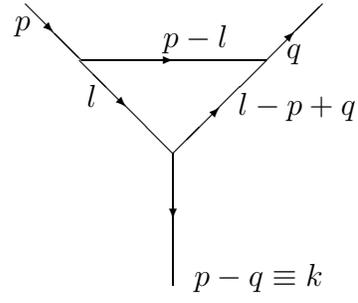
\begin{figure}
\begin{center}
\begin{picture}(10000,15000)
\drawline\fermion[\SE\REG](1000,12000)[3000]
\drawarrow[\LDIR\ATTIP](\pmidx,\pmidy)
\global\advance\fermionfrontx by -400
\global\advance\fermionfronty by -1000
\put(\fermionfrontx,\fermionfronty){$p$}
\drawline\fermion[\E\REG](\fermionbackx,\fermionbacky)[7000]
\drawarrow[\LDIR\ATTIP](\pmidx,\pmidy)
\global\advance\pmidy by 500
\global\advance\pmidx by -400
\put(\pmidx,\pmidy){$p-l$}
\drawline\fermion[\NE\REG](\fermionbackx,\fermionbacky)[3000]
\drawarrow[\LDIR\ATTIP](\pmidx,\pmidy)
\global\advance\pmidx by -300
\global\advance\pmidy by -1000
\put(\pmidx,\pmidy){$q$}
\drawline\fermion[\SW\REG](\fermionfrontx,\fermionfronty)[5000]
\drawarrow[\NE\ATTIP](\pmidx,\pmidy)
\global\advance\pmidx by 700
\global\advance\pmidy by -300
\put(\pmidx,\pmidy){$l-p+q$}
\drawline\fermion[\S\REG](\fermionbackx,\fermionbacky)[5000]
\drawarrow[\LDIR\ATTIP](\pmidx,\pmidy)
\global\advance\fermionbackx by 800
\put(\fermionbackx,\fermionbacky){$p-q\equiv k$}
\drawline\fermion[\NW\REG](\fermionfrontx,\fermionfronty)[5000]
\drawarrow[\SE\ATTIP](\pmidx,\pmidy)
\global\advance\pmidx by -1500
\put(\pmidx,\pmidy){$l$}
\end{picture}
\end{center}
\caption{momentum routing in the Feynman diagrams}
\label{m-route}
\end{figure}

\begin{thebibliography}{99} 
\bibitem{emkw} D. Evens, J.W. Moffat, G. Kleppe and R.P. Woodard,
Phys. Rev. {\bf D43}, 499 (1991).

\bibitem{kw} G. Kleppe and R.P. Woodard, Nucl. Phys. {\bf B388},
81 (1992).

\bibitem{kws} G. Kleppe and R.P. Woodard, Ann. Phys. (N.Y.), {\bf 221},
106 (1993).

\bibitem{cdm} M.A. Clayton, L. Demopoulos and J.W. Moffat, Int. J. Mod. Phys.,
{\bf A9}, 4549 (1994).

\bibitem{par} J. Paris, Nucl. Phys., {\bf B450}, 357 (1995).

\bibitem{brod} S.J. Brodsky et al., Phys. Rev., {\bf D39}, 2797 (1989).

\bibitem{ch-li} T.P. Cheng and L.F. Li, {\em Gauge Theory of Elementary
Particle Physics}, Oxford University Press, (1984).

\bibitem{nam} See, for example, K. Namsrai, {\em Nonlocal Quantum
Field Theory and Stochastic Quantum Mechanics}, D. Reidel Publishing
Company, (1986) and references therein.

\bibitem{pec} For a discussion on composite models see R.D. Peccei in
{\em Gauge Theories of the Eighties}, Lecture Notes in Physics, vol. 181,
Spriger-Verlag (1983), Edited by R. Raitio and J. Lindfers.

\bibitem{fp} F.J.M. Farley and E. Picasso, {\em The muon g-2 Experiment},
in Advances Series on Directions in High Energy Physics, vol. 7, World
Scientific Publishing Co, 1990, Edited by T. Kinoshita.

\bibitem{kin} T. Kinoshita and W.J. Marciano, {\em Theory of Muon Anomalous
Magnetic Moment} , in Advanced Series on Directions in High Energy Physics,
vol.7, World Scientific Publishing Company, 1990, Edited by T. Kinoshita.

\bibitem{vwh} V.W.Hughes in Frontiers in High Energy Physics,Universal
Academy Press Tokyo,1992,Edited by T.hasegawa(p.717)

\bibitem{holl} W.Hollik,{\em Review of electroweak theory},hep-ph/9610457.
\bibitem{mery} P.Mery,S.E Moubarik,M.Perottet,F.M Renard,Z.Phys.C,
{\bf C46},229(1990).
\bibitem{Car} M.Carena,G.F Guidice,C.E.M Wagner,hep-ph(9610233).
\bibitem{arzt} C.Arzt,M.B.Einhorn,J.Wudka,Phys.Rev{\bf D49},1370(1994). 
\end{thebibliography}
\end{document}